\documentclass[aps,prb,twocolumn,showpacs,floatfix,preprintnumbers,superscriptaddress,amsmath,amssymb,footinbib]{revtex4-1}

\usepackage{graphicx}
\usepackage{dcolumn}
\usepackage{bm}

\begin{document}

\title{Ferromagnetism of Pd(001) substrate induced by Antiferromagnetic CoO}

\author{S. K. Saha}
\affiliation{Max-Planck-Institut f\"{u}r Mikrostrukturphysik,
06120 Halle, Germany}
\author{V. S. Stepanyuk} \affiliation{Max-Planck-Institut f\"{u}r Mikrostrukturphysik, 06120 Halle, Germany}
\author{J. Kirschner}
\affiliation{Max-Planck-Institut f\"{u}r Mikrostrukturphysik, 06120 Halle, Germany}

\date{\today}

\begin{abstract}
Our  first-principles study has revealed
unexpected spin polarization of the Pd(001) substrate in contact
with antiferromagnetic CoO overlayers. 
We give an evidence that the ferromagnetism of Pd is caused by the zigzag positions of Co atoms with respect to the Pd interface, resulted from the lattice-mismatch driven structural relaxation. Thanks to the itinerant nature of its 4d electrons, we see that the ferromagnetic properties of Pd are highly sensitive to the local environment and can be enhanced further by increasing the thickness of CoO overlayer film or/and by applying an additional uniaxial pressure along c-axis exerted externally on the bottom layers of the Pd substrate. Our finding provides new functionality for the interfacial moments of the CoO/Pd system, which can be accessed experimentally, e.g., by the magneto-optical Kerr effect (MOKE) or/and by element-resolved X-ray magnetic circular dichroism (XMCD) measurement.
\end{abstract}


\maketitle

Magnetism at interface is a fascinating field of research
\cite{sri1, sri2, sri3, sri4}. The reduced dimensionality and the
substrate induced effects such as epitaxial strain, charge
transfer and electronic hybridization, often exhibit extremely
fascinating effects that are of fundamental interest and
technological importance, largely for magnetic data storage
and nanoelectronics \cite{sri5,sri6}.

In particular, the interface effects and the ability to precisely
engineer them can enable making a nonmagnetic material magnetic
and switching its magnetism on and off , which are the holy grails
of spintronics. The fact that the magnetism emerges from a
nonmagnet being in touch with ferromagnet is expected \cite{sri1},
but the same arising from a nonmagnet by contact with an
antiferromagnet is puzzling since one thinks of antiferromagnets
as magnetically neutral.

\begin{figure}[b]
\centering \vspace*{-0.3cm}\includegraphics[scale=0.35]
{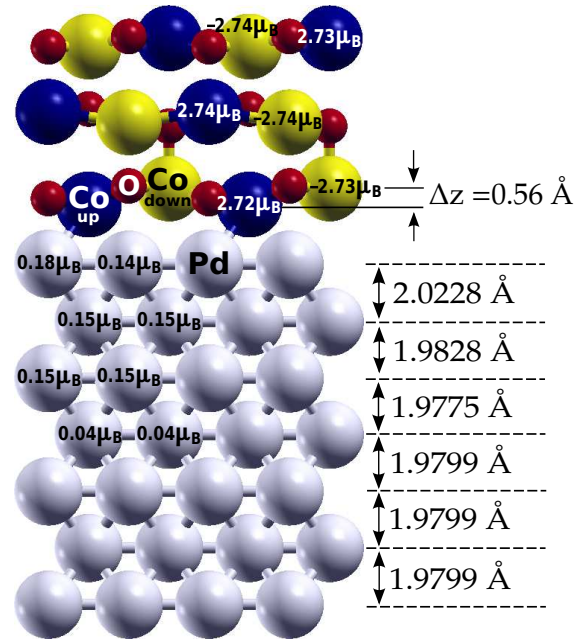} \vspace*{-0.2cm} \caption{(color online) Pictorial
side view of the calculated structure for 3\,ML of CoO on Pd(001)
substrate. Our structural relaxation, which allows selectively
dynamics along z-axis, suggests that, at the interface overlayer,
the lattice misfit-induced strain brings Co-up much closer to Pd
substrate than Co-down. The interlayer distances, magnetic moments
of Co as well as induced magnetic moments of Pd are
specified.}\label{Fig.1}
\end{figure}

Already, upon placing antiferromagnet on a ferromagnet the effects
of enhanced coercivity and exchange bias have been reported
\cite{sri2, sri3, sri4}. Now, when antiferromagnet is interfaced
with a nonmagnet, it is highly exciting to see whether the same
behavior can be met, i.e., whether exchange bias can be observed
in the interface between antiferromagnet and originally
nonmagnetic material. From application point of view, it would be
extremely interesting to switch on magnetism and simultaneously
pin the magnetization through the exchange bias to an
antiferromagnetic layer.

\begin{figure*}[!]
\centering \vspace*{-0.3cm}\includegraphics[scale=0.4] {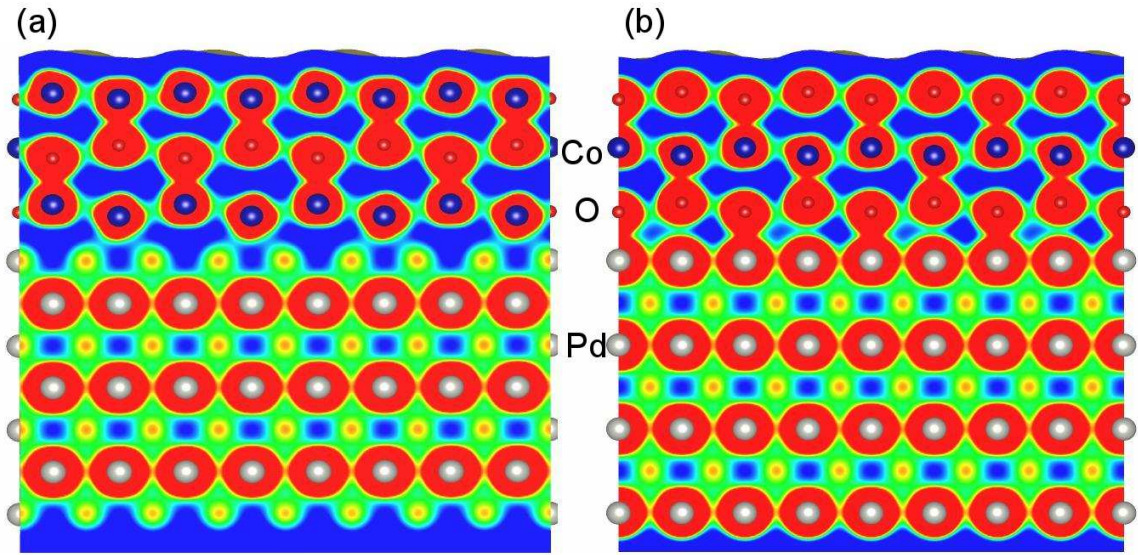}
\vspace*{-0.2cm} \caption{(color online)
 (a) The color pattern of the calculated valence charge density distribution
for 3 ML of CoO on Pd(001) with a cut by [100] plane through Co atoms at interface in order to
show how Co atoms couple with Pd atoms at interface, and (b) the same with a cut by [100] plane
through O atoms at interface in order to show how O atoms couple with Pd atoms at interface.
Deep blue balls (dark and big) represent Co atoms, deep red balls (light and small) represent O
atoms and grey balls represent Pd atoms. Besides balls, other blue (dark) represents low valence
charge while other red (light) represents a high concentration of valence charge.
}\label{Fig.2}
\end{figure*}

Since palladium is an element very close to fulfill the so-called
Stoner criterion of ferromagnetism \cite{sri7}, it makes
nonmagnetic Pd a good candidate for such studies. Much attention
has been paid to the magnetic properties of Pd: it was
shown that the effect of ferromagnetism in Pd
can be achieved by reduced dimensionality
\cite{Sam2003,Shin2003} and/or by quantum well states
formed in ultrathin Pd films \cite{Hon2007}. It is also well known
that the neighboring \emph{3d}-ferromagnet can spin-polarize a
non-magnetic metal like Pd \cite{Ful1995}. Many different
experimental techniques and theoretical studies have been applied
to investigate various combinations of ferromagnet/Pd systems not
only to understand the mechanism behind such spin-polarization
effect, but also for possible applications
\cite{Blu1988,Che2004,Lu2005}. Especially, the magneto-optical
Kerr effect (MOKE) measurements provide valuable information about
such systems because the magneto-optical response from the
ferromagnet/Pd interface is large
\cite{Lu2005,Cho2001,Oh2001,Lu2006}. 

The polarization of Pd by AFM was considered for the first time in
NiO/Pd multilayers \cite{Man1998}. Manago et al. \cite{Man1998,
Man1999a, Man1999b} suggested that the ferromagnetic properties of
such multilayers are related to the induced magnetic moment of Pd
of 0.59\,$\mu_{B}$ per interface Pd atom, which persists over a
dozen monolayers (ML) away from the Pd/NiO interface. In
comparison, the theoretical calculation for FM/Pd systems shows
that both the magnetic moments and the number of polarized Pd
atomic layers are much smaller \cite{Blu1989}. Later on, Hoffmann
et al. \cite{Hof2001} investigated (NiO/Pd)$_{N}$ systems, in a
form of bilayers ($N$=1) and multilayers, grown on different
substrates by applying polarized neutron reflectometry. In each of
the cases, no effect of ferromagnetic Pd was observed. It should
be emphasized, however, that both Manago et al. and Hoffmann et
al. studied the polarization of Pd in systems consisting of thin
Pd films, and with the NiO/Pd sequence repeated many times. In
such cases, the structural parameters (crystallographic
orientation, lattice constant, quality of the film) cannot be
represented by the model system. In particular, such structural
modifications can be crucial for the spin-polarization effect, as
it will be discussed in this letter.

Up to now there are no theoretical investigations showing how the
AFM-film can drive ferromagnetism when it is in direct contact
with a non-ferromagnetic metal like Pd. Motivated by these
considerations, we show for the first time how the AFM thin film
(of CoO) induces ferromagnetism in a nonmagnetic substrate [of
Pd(001)]. Our finding provides new functionality for the
interfacial moments of the AFM(CoO)/FM(Pd) system which can be
accessed experimentally, e.g., by MOKE or/and by element-resolved X-ray magnetic circular dichroism (XMCD) measurement.

The crystallographic and spin structure of antiferromagnetic CoO
can be described as follows: each Co atom has a spin that is
antiparallel to the next Co atom (see Fig.\ref{Fig.1}), where
AFM-Pd could be expected rather than FM-Pd. Therefore, to
understand the ferromagnetic behavior of Pd, we study the
CoO/Pd(001) system theoretically, using the VASP \cite{vasp}
implementation of DFT, with projector augmented wave (PAW)
potentials \cite{paw}. We adopt the exchange-correlation
functional of Perdew-Wang (PW91) \cite{pw91} for the generalized
gradient approximation (GGA) and an effective Hubbard $U$ \cite{sks} of
6.1\,eV for Co. Kohn-Sham wave functions are represented using a
plane-wave basis truncated at an energy cutoff of 40\,Ry and the
Brillouin zone integrations are done on a uniform Monkhorst-Pack
\cite{mpg} grid of 19$\times$19$\times$1. We use supercell
geometry with a vacuum of 10\AA ~in the z direction to ensure
negligible interaction between its periodic images. Selected
structural relaxation is carried out so as to minimize the forces
acting on each of the atoms using a conjugate-gradient algorithm.

The Pd(001) substrate is simulated by a 7-layer slab with the
interlayer distances of the three bottom-most layers keeping fixed
in its bulk GGA-optimized value of 1.9799\AA ~(see
Fig.\ref{Fig.1}). Pd(001) has a primitive square net with an
in-plane lattice spacing of 2.80\AA, which does not match to that
of CoO (3.01\AA). This large in-plane lattice mismatch causes an
elastic strain. This strain, as well as the presence of dissimilar
neighbor such as those found at the interface and the absence of
some neighbors such as those found at the surface, leads to
structural relaxations resulting in considerable structural
inhomogeneities. Our structural relaxation, which selectively
allows dynamics along z-axis suggests that the lateral registry
between substrate and overlayer, where O atoms sit on top of the
outermost Pd atoms and Co in fourfold hollow sites, is
energetically favorable. This relaxation results in substantially
distorted bond angles, different bond lengths and interlayer
distortion as evident from Fig.1 and Fig.2. At the interface
overlayer, this interlayer distortion brings Co-up much closer to
Pd substrate (0.56\AA) than the Co-down atoms.

\begin{figure}[tbp]
\centering \vspace*{-0.3cm}\includegraphics[scale=0.7] {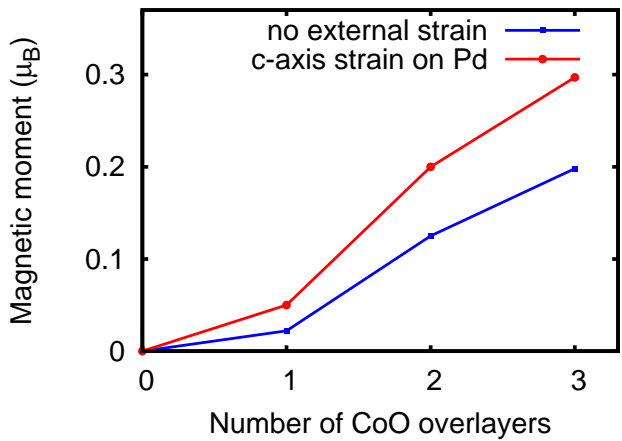}
\vspace*{-0.2cm} \caption{(color online)  
The calculated induced magnetic moment on Pd interface atoms for
(001) substrate as a function of the number of CoO overlayers is
plotted. The blue curve (darker) is obtained when
the Pd(001) substrate is simulated by a 7-layer slab with the
interlayer distances of the three bottom-most layers keeping fixed
in its bulk GGA-optimized value of 1.9799\AA, and the red curve
(lighter) is for the same system but an additional uniaxial strain
(about 1.7\%) along c-axis is applied externally on the three
bottom-most layers of the Pd substrate.}\label{Fig.3}
\end{figure}

Such structural inhomogeneities evidently influence both the local
and long-range magnetic ordering of the CoO overlayers, resulting
in a complex magnetic pattern. For example, an uncompensated
magnetic configuration near the interface overlayer results from
these structural inhomogeneities. When three overlayers of CoO are
placed on the Pd substrate, the magnetic moments of Co become
2.72$\mu_{\rm B}$ (site up) and -2.73 $\mu_{\rm B}$ (site down) at
the interface overlayer ($I$). For the layers $I$+1 and $I$+2,
these values are 2.74 $\mu_{\rm B}$ , $-$2.74 $\mu_{\rm B}$ and
2.73 $\mu_{\rm B}$, $-$2.74 $\mu_{\rm B}$, respectively. The
magnetic moment is not completely compensated because the relaxed
positions of the Co atoms are no longer equivalent by symmetry and
show a zigzag configuration with respect to the Pd surface.
The zigzag atom configuration causes spin uncompensation, which in fact
induces magnetic moments in Pd layers near the interface, and
turns an otherwise nonmagnetic Pd substrate into a ferromagnet.

Note, that, in the case of Ir(001) substrate, the CoO(111)
monolayer, where the neighboring Co atoms have, at least
partially, opposite magnetic moments, the polarization of the
nearby Ir atoms is small \cite{Mittendorfer}. 
The
presence of ferromagnetism in Pd/CoO system is a key  result of our
study which can be attributed to a combination of several reasons.
First of all, Pd is very close to fulfill the so-called Stoner
criterion of ferromagnetism due to a high-density of electronic
states near its Fermi energy. Secondly, metallic palladium itself
is not ferromagnetically ordered, though its magnetic
susceptibility is very large due to a high density of electronic
states near its Fermi energy. Consequently, such spin induction
can stabilize the ferromagnetic order in Pd. Furthermore, in Pd
the large itinerant nature of its \emph{4d} electrons (its
\emph{4d} electrons are much more delocalized than in Ir and Rh)
causes a nearly-itinerant-electron like parabolic band structure \cite{saha}.
This facilitates itinerant band ferromagnetism in Pd and makes its
ferromagnetic properties a sensitive function of local
environment. As a result, by changing the thickness of CoO
overlayer or by reducing interlayer distance a bit for the three
bottom-most Pd layers (which can be achieved by applying an
external uniaxial pressure along c-axis on the bottom of the Pd
substrate), its magnetic moment can be enhanced further, as shown
in Fig.\ref{Fig.3}.

Finally, the spin-polarized Pd behaves like a ferromagnetic film
in direct contact with an antiferromagnetic CoO layer. Thus, one
can expect a presence of the exchange bias effect (a shift of the
loops along the field axis by $H_{EB}$) and an increase of the
coercivity field which is considered as an indication of coupling
\cite{Nog2005}. Note, however, that the AFM/FM interaction is very
sensitive to interface conditions, and most of the parameters
involved in such interaction (such as structural imperfection,
domain formation and/or spin configuration) are difficult to
control.

In conclusion, we demonstrate how the proximity
of an AFM thin film (of CoO) drives magnetization to a nonmagnetic
metal substrate [of Pd(001)], using a 
first-principles density functional theoretical investigation. 
We show that the
ferromagnetism of Pd results from the zigzag positions of Co atoms
caused by the strain-relief mechanism near the interface. 
We expect that this theoretical study will be
of value in the context of experimental work, especially in MOKE and XMCD, 
 to develop a better understanding of novel functionalities for 
the interfacial moments
of the CoO/Pd(001) system, which furthermore
could incite futuristic device applications based on the exciting ability 
to control the functionality for such interfacial moments.

We are thankful to M. Przybylski for helpful discussion.
SKS and VSS gratefully acknowledge the support by
the German joint research network Sonderforschungsbereich 762
“Functionality of oxidic interfaces” of the Deutsche
Forschungsgemeinschaft.

\end{document}